\documentclass[aps,pre,twocolumn,showpacs,floatfix,superscriptaddress]{revtex4}
\usepackage{graphicx}

\begin{document}
\title{Nonlinear supratransmission and bistability in the Fermi-Pasta-Ulam model}

\author{Ramaz Khomeriki} 
\email{khomeriki@hotmail.com}
\affiliation{Tbilisi State University, 3 Chavchavadze Avenue, 
Tbilisi 380028, Republic of Georgia}

\author{Stefano Lepri}
\email{stefano.lepri@unifi.it}
\affiliation{Istituto Nazionale per la Fisica della Materia, 
Unit\`a di Firenze, via G. Sansone 1, 50019 Sesto Fiorentino, Italy}

\author{Stefano Ruffo} 
\email{ruffo@avanzi.de.unifi.it}
\affiliation{Dipartimento di Energetica ``S. Stecco" and CSDC, 
Universit\'a di Firenze, Via S. Marta, 3, I-50139 Firenze, Italy}
\affiliation{Istituto Nazionale di Fisica Nucleare, Sezione di Firenze, 
via G. Sansone 1, 50019 Sesto Fiorentino, Italy}
\affiliation{Istituto Nazionale per la Fisica della Materia, 
Unit\`a di Firenze, via G. Sansone 1, 50019 Sesto Fiorentino, Italy}

\date{\today}

\begin{abstract}

The recently discovered phenomenon of nonlinear {\it supratransmission} 
consists in a sudden increase of the  amplitude of a transmitted wave triggered 
by the excitation of nonlinear localized modes of the medium. We examine this process 
for the Fermi-Pasta-Ulam chain, sinusoidally driven at one edge and damped
at the other. The 
supratransmission regime occurs for driving frequencies above the upper band-edge 
and originates from direct moving discrete breather creation. We derive approximate analytical 
estimates of the supratransmission threshold, which are in excellent agreement with
numerics. When analysing the long-time behavior, we discover that, below the
supratransmission threshold, a {\it conducting} stationary state 
coexists with the {\it insulating} one. We explain the {\it bistable} nature of the
energy flux in terms of the excitation of quasi-harmonic extended waves.
This leads to the analytical calculation of a {\it lower-transmission} threshold
which is also in reasonable agreement with numerical experiments.


\end{abstract}

\maketitle

\section{Introduction}

In a recent series of interesting papers J. Leon and coworkers~\cite{leon,leon2,leon3,leon4}
discovered that nonlinear chains driven at a boundary can propagate energy in the 
forbidden band gap. Numerical experiments were performed for harmonic driving,
and the semi-infinite chain idealization was simulated by adding damping on 
the boundary opposite to 
driving. In this case, energy transmission occurs above a well defined (frequency dependent)
critical amplitude. This phenomenon has been called {\it nonlinear supratransmission}
by the authors, and is characterized by the propagation of nonlinear localized modes
(gap solitons) inside the bulk. Several models have been considered: sine-Gordon 
and Klein-Gordon~\cite{leon}, double sine-Gordon and Josephson transmission lines~\cite{leon2}, 
Bragg media~\cite{leon4}, and an experimental realization has been proposed for a mechanical 
system of coupled pendula~\cite{leon2}. The generic features of the supratransmission
instability have been described in terms of an evanescent wave destabilization~\cite{leon3}.
Moreover, the same process has been described in Ref.~\cite{ramaz1}
for the discrete nonlinear Schr\"odinger equation, suggesting
an experimental application to optical waveguide arrays.   

In this paper we show that the supratransmission phenomenon is present for Fermi-Pasta-Ulam (FPU)
nonlinear chains~\cite{fermi}. At variance with all previously considered
cases, the harmonic driving frequency must lie above the phonon band, since the FPU interparticle 
potential is translationally invariant and, hence, a forbidden lower band does not
exist (the phonon spectrum begins at zero frequency). This entails that the nonlinear modes 
which propagate in the bulk are moving {\it discrete breathers}~\cite{flach}. Exact static 
discrete breathers profiles have been presented in the literature, but here we use approximate analytic expressions for both
the low-amplitude solitonic case and for the large amplitude situation~\cite{kosprl}.
This allows to perform a study of the instability at the boundary and a detailed 
analysis of the process which leads to the birth and the propagation of the discrete breather. 
By using these approximate solutions, we are able to provide analytic expressions 
for the supratransmission critical amplitudes as
a function of the forcing frequencies, which are then successfully compared with 
numerically determined values.

Besides that, we analyse the long-time behavior of the system, studying the formation
of a stationary state with a given energy flux across the chain. The {\it order parameter}
of the transition from the {\it insulating} to the {\it conducting} state is, indeed, the average energy flux, which displays
a jump at the supratrasmission threshold (which could then be thought as a sort
of non-equilibrium first-order transition). We discover that lowering the amplitude 
below the threshold, after the stationary state is established, does not interrupt
trasmission: the {\it conducting} state survives even at smaller amplitudes and
coexists with the {\it insulating} state (a sort of bistability is present in the 
system). By further reducing the amplitude, a threshold appears
below which the energy flux vanishes without any apparent discontinuity (here
we have a sort of second-order transition): we develop a theoretical analysis of this
new threshold phenomenon, which was absent in previous studies.   
 
The paper is organized as follows. In Section II we introduce the model and the
equations of motion. Section III deals with the calculation of the energy flux in
the quasi-linear approximation. Section IV illustrates all analytic and numerical results
concerning the determination of the supratransmission threshold. Section V is devoted
to the characterization of the stationary states and of their bistability. Section VI contains some conclusions.
In the Appendix we report, for completeness, a calculation of the nonlinear phonon 
dispersion relation.

\section{The model}

We consider the Fermi-Pasta-Ulam (FPU) chain~\cite{fermi}, which is an 
extremely well studied nonlinear lattice for which a large class of quasi-harmonic and 
localized solutions is known. 
The equations of motion for the so-called $\beta$-FPU chain (interparticle potential 
with a quadratic and a quartic term) are
\begin{equation}
\ddot u_n =u_{n+1}+u_{n-1}-2u_n+(u_{n+1}-u_n)^3+(u_{n-1}-u_n)^3~, 
\label{1}
\end{equation}
where $u_n$ stands for the displacement of $n$-th site in dimensionless units
($n=0,1,2\ldots, N$).
All force parameters have been chosen equal to unity for computational convenience.

To simulate the effect of an impinging wave we impose the boundary condition
\begin{equation}
u_0(t) \;=\; A \cos \omega t \quad .
\label{drive}
\end{equation}
Free boundary conditions are enforced on the other side of the chain. 

In order to be able to observe a stationary state in the {\it conducting} regime we need to
steadily remove the energy injected in the lattice by the driving force.  Thus, we
damp a certain number of the rightmost sites (typically 10\% of the total) by
adding a viscous term $-\gamma \dot u_n$ to their equations of motion.  A convenient
indicator to look at is the averaged energy flux $j=\sum_n j_n/N$, where the local
flux $j_n$ is given by the following formula~\cite{rep}
\begin{equation}
j_n=\frac12 (\dot u_n+\dot u_{n+1})\left[u_{n+1}-u_n+(u_{n+1}-u_{n})^3\right]. 
\label{flux}
\end{equation} 
Time averages of this quantity are taken in order to characterize
the insulating (zero flux)/conducting (non zero flux) state of the system.

\section{In-band driving: nonlinear phonons}

For illustration, we first discuss the case when the driving frequency is located
inside the phonon band. Although trivial, this issue is of importance to better
appreciate the fully nonlinear features described later on.

Under the effect of the driving (\ref{drive}), we can look for extended
quasi-harmonic solutions (nonlinear phonons) of the form
\begin{equation}
u_n \; \; =A\cos(kn-\omega t) \quad .
\label{0}
\end{equation}
We consider the semi--infinite chain, so that $k$ varies continously between $0$
and $2\pi$. The nonlinear dispersion relation can be  found in the
rotating wave approximation (see e.g. Ref.~\cite{rwa}).
Neglecting higher--order harmonics (see the Appendix for details) it reads
\begin{equation}
\omega_0^2(k,A) = 2(1-\cos k)+3(1-\cos k)^2 A^2~. 
\label{2}
\end{equation}
Thus the nonlinear phonon frequencies range from 0 to the upper band--edge 
$\omega_0(\pi,A)\geq 2$.

\begin{figure}[h]
\begin{center}\leavevmode
\includegraphics[width=0.8\linewidth,clip]{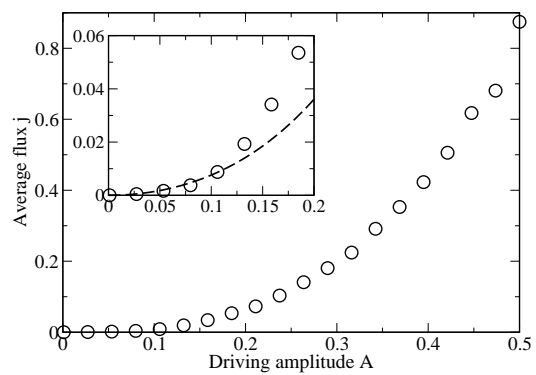}
\end{center}
\caption{Average energy flux vs. 
driving amplitude for in-band forcing, $\omega=1.8$, $\gamma=5$.
Data have been averaged over $10^5$ periods of the driving. The
inset is an enlargment of the small-amplitude region and the dashed
line is the single nonlinear phonon approximation (\protect\ref{jphon}).
}
\label{inband}
\end{figure}

If we simply assume that only the resonating phonons whose wavenumbers satisfy
the condition 
\begin{equation}
\omega=\omega_0(k,A) 
\end{equation}
are excited, we can easily estimate the energy flux.
Neglecting, for simplicity, the nonlinear force terms in the definition of the
flux (\ref{flux}), we have
\begin{equation}
j = \frac12 v(k,A) \, \omega^2 A^2 \quad,
\label{jphon}
\end{equation}
where $v$ is the group velocity as derived from dispersion  relation
(\ref{2}).  This simple result is in very good agreement with simulations,
at least for small enough amplitudes (see Fig.~\ref{inband}). 
For $A > 0.15$  the measured flux is larger than the estimate (\ref{jphon}), indicating
that something  more complicated occurs in the bulk (possibly, a multiphonon
transmission) and that higher-order nonlinear terms must be taken into account.

\section{Out-band driving: supratransmission}

Let us now turn to the more interesting case in which the driving frequency lies
outside the phonon band, $\omega>\omega_0(\pi,0)=2$. In a first series of numerical
experiments we have initialized the chain at rest and switched on the driving at time
$t=0$. To avoid the formation of sudden shocks~\cite{shocks}, we have chosen to increase smoothly
the amplitude from 0 to the constant value $A$ at a constant rate, i.e. 
\begin{equation}
u_0=A\cos(\omega t)\left[1-e^{-t/\tau_1}\right]~,
\label{dr1} 
\end{equation} 
where typically we set $\tau_1=10$.
\begin{figure}[b]
\begin{center}\leavevmode
\includegraphics[width=0.8\linewidth,clip]{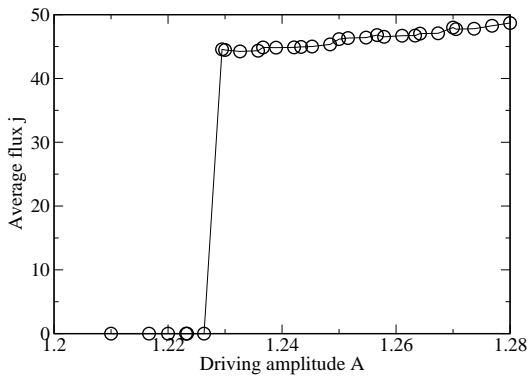}
\end{center}
\caption{Average energy flux vs. 
driving amplitude for out--band forcing, $\omega=3.5$, $\gamma=5$.
Data have been averaged over $2\,10^5$ periods of 
the driving for a chain of $N=512$ particles.}
\label{outband}
\end{figure}

At variance with the case of in--band forcing, we observe a sharp increase of the flux
at a given threshold amplitude of the driving, see Fig.~\ref{outband}.
This  phenomenon has been denoted as {\it nonlinear supratransmission}~\cite{leon} to 
emphasize the role played by nonlinear localized excitations in triggering the energy flux.
 
This situation should be compared with the one of in--band driving, shown in 
Fig.~\ref{inband}, where no threshold for conduction exists and the flux
increases continuously from zero (more or less quadratically
in the amplitude). 
Indeed, the main conclusion that can be drawn from the previous section is that there
cannot be any amplitude threshold for energy transmission in the case of
in-band forcing. Moreover, although at the upper band edge the flux vanishes, since
it is proportional to the group velocity (see formula (\ref{jphon})), it is 
straightforward to prove that it goes to zero with the square root of the distance to the band
edge frequency. Hence, the sudden jump we observe in the out-band case cannot 
be explained by any sort of quasi-linear approximation. 

In the following we investigate the physical origin of nonlinear supratransmission,
distinguishing the cases of small and large amplitudes.
\begin{figure}[b]
\begin{center}\leavevmode
\includegraphics[width=0.7\linewidth,angle=-90]{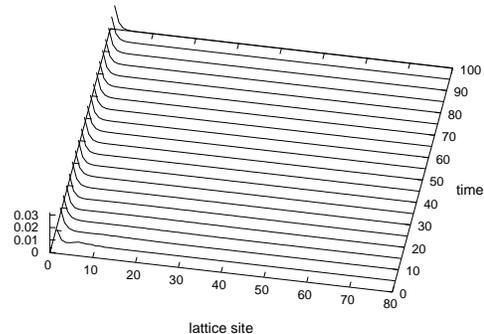}
\includegraphics[width=0.7\linewidth,angle=-90]{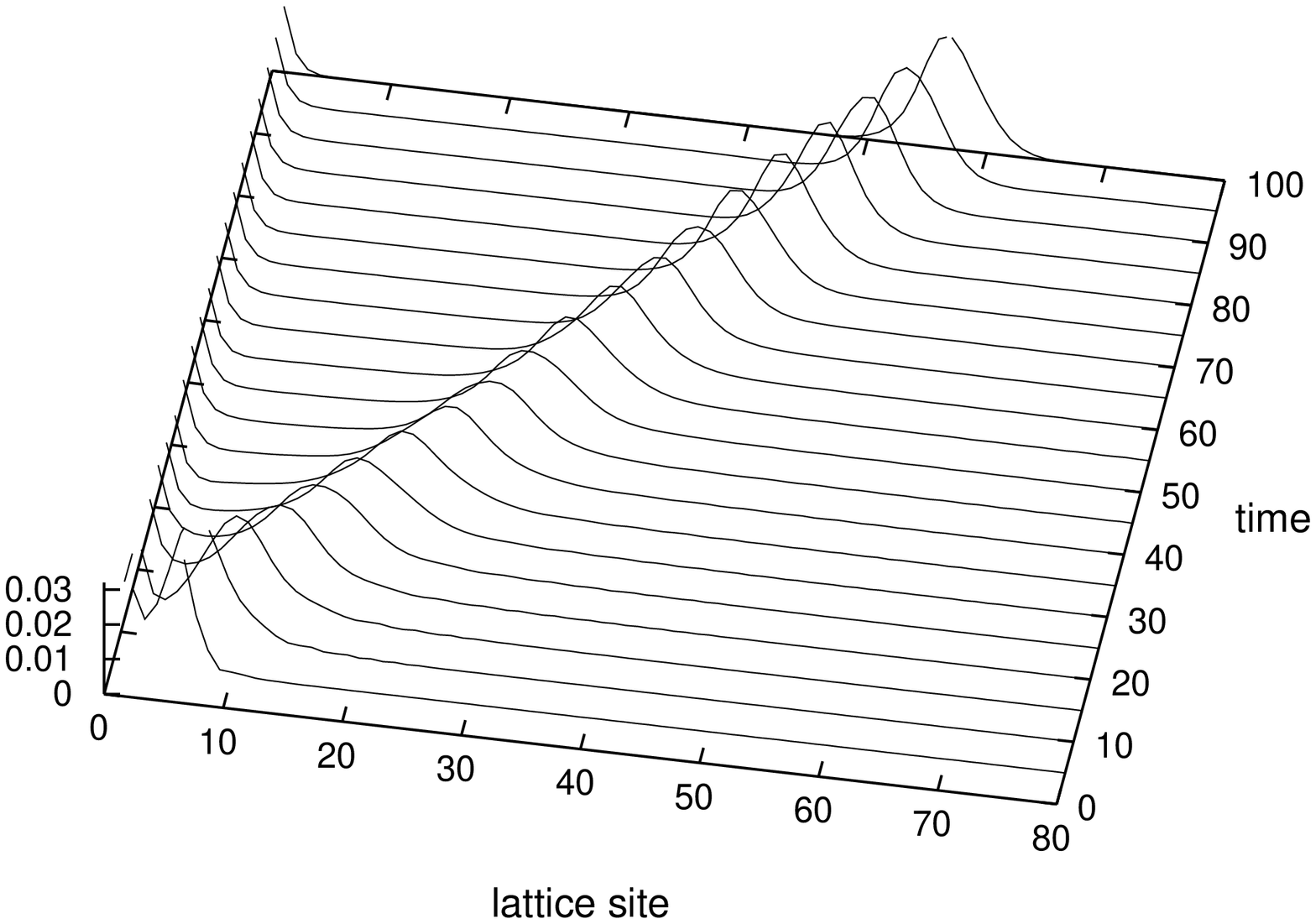}
\caption{Snapshot of the local energy 
below the supratransmission threshold $A=0.15 < A_{th}$ for 
$\omega =2.1$ $\gamma=10$. The initial condition is an envelope soliton
(\ref{sol}) with $x_0=-1.8$ (above) and $x_0=+1.8$ (below).
}
\label{2brea}
\end{center}
\end{figure}

\subsection{Small amplitudes}

When the driving frequency is only slightly above the band ($0<\omega-2 \ll 1$),
one can resort to the continuum envelope approximation. Since we expect the 
zone--boundary mode $k=\pi$ to play a major role, we let
\begin{equation}
u_n \;=\;  (-1)^n \frac12 \left[ \psi_n\, e^{i\omega t} + \psi_n^*\, 
e^{-i\omega t}\right]~.
\end{equation}
In the rotating wave approximation~\cite{rwa} and for slowly varying 
$\psi_n$ one obtains from the FPU lattice equations the nonlinear 
Schr\"odinger equation ($\psi_n \longrightarrow \psi(x,t)$)~\cite{scott}
\begin{equation}
2i\omega \dot\psi \;=\; (\omega^2 - 4) \psi -\psi_{xx} -12 \psi|\psi|^2~,
\label{nls}
\end{equation} 
with the boundary condition $\psi(0,t)=A$.

The well-known {\it static} single--soliton solution of Eq.~(\ref{nls}) corresponds to the 
family of envelope solitons (low-amplitude discrete breathers)
\begin{equation}
u_n \;=\; a (-1)^n \cos(\omega t) \, {{\rm sech}\left[\sqrt{6} (n-x_0)a\right]},
\label{sol}
\end{equation}
with amplitude $a=\sqrt{(\omega^2-4)/6}$. The maximum of the soliton shape is 
fixed by the boundary condition to be
\begin{equation} 
x_0 \;=\; \pm {{\rm acosh}(a/A) \over a\sqrt{6}}~.
\label{x0}
\end{equation}
In this approximation we have two possible solutions: one with the maximum outside
the chain, which is purely decaying inside the chain (minus sign in (\ref{x0})), and another 
with the maximum located within the chain (plus sign in (\ref{x0})).
Overcoming the supratrasmission threshold corresponds to the disappearence of both solutions. 
Indeed, when the driving amplitude reaches the critical value $A_{th}$, given by 
\begin{equation}
\omega^2=4+6A_{th}^2~,
\label{5}
\end{equation} 
solution (\ref{sol}) ceases to exists. 

\begin{figure}[b]
\begin{center}\leavevmode
\includegraphics[width=0.7\linewidth,angle=-90]{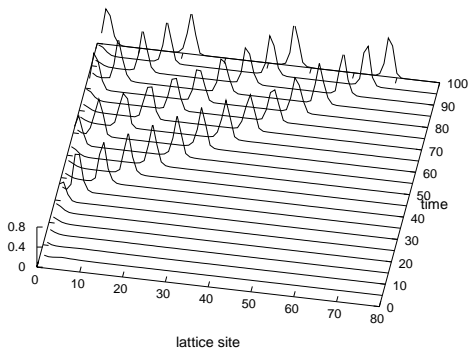}
\caption{Snapshot of the local energy 
at the transmission threshold $A=0.253 \approx A_{th}$ for $\omega =2.1$  
$\gamma=10$. The initial condition is the envelope soliton (\ref{sol})
with $x_0\approx 0$.}
\label{transm}
\end{center}
\end{figure}

We have investigated this issue by simulating the 
lattice dynamics with the initial conditions given by Eqs.~(\ref{sol}) and (\ref{x0}).
The evolution of the local energy
\begin{equation} 
\label{localenergy}
e_n \;=\; {\dot u^2_n \over 2}+\frac12 \left[V(u_{n+1}-u_{n})+V(u_{n}-u_{n-1})\right]
\end{equation}
with $V(x)=x^2/2+x^4/4$,  
is shown in Fig.~\ref{2brea}. The solution with the maximum outside the chain 
(upper figure) stabilizes after the emission of a small amount of radiation (generated by
the fact that we have used an approximate solution). On the contrary, the other 
solution (lower figure) slowly moves towards the right and, eventually, leaves 
a localized boundary soliton behind. The release of energy to the 
chain is non stationary and does not lead to a conducting state. 

The scenario drastically changes at the supratransmission amplitude $A_{th}$. The 
chain starts to conduct: a train of {\it travelling} envelope solitons is emitted from
the left boundary (see Fig.~\ref{transm}). Here we should emphasize that 
the envelop soliton solution (\ref{sol}), which is characterized by the $k=\pi$ carrier 
wave--number, has a zero group velocity. Thus, transmission cannot be realized by such
envelope solitons. 
Instead, transmission starts when the driving frequency resonates with the 
frequency of the envelope soliton with carrier wave--number $k=\pi(N-2)/N$, next 
to the $\pi$-mode. However, as far as we consider a large number of oscillators ($N=500$), 
we can still use expression (\ref{5}) for the $\pi$-mode frequency.

\subsection{Large amplitudes}

The above soliton solution is valid in the continuum envelope limit, and is therefore less and 
less accurate as its amplitude increases. Indeed, if the weakly nonlinear condition
is violated, the width of the envelope soliton becomes comparable with 
lattice spacing and, thus, one cannot use the continuum envelope approach. 
Fortunately, besides the slowly varying envelope soliton solution (\ref{sol}), 
an analytic approximate expression exists for large amplitude static 
discrete breather solutions, which is obtained from an exact extended plane wave 
solution with ``magic" wave--number $2\pi/3$~\cite{kosprl}
\begin{equation}
u_n=a(-1)^n \cos\left[\omega_B(a) t\right]\cos\left(\frac{\pi}{3}n\pm x_0\right)~, 
\label{breather}
\end{equation}
if $\left|(\pi n/3)\pm x_0\right|<\pi/2$ and $u_n=0$ otherwise. 

\begin{figure}[b]
\begin{center}\leavevmode
\includegraphics[width=0.85\linewidth]{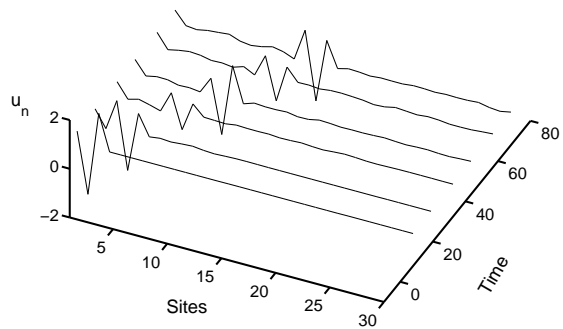}
\caption{Snapshot of particle displacements $u_n$ below the supratransmission 
threshold for a driving frequency $\omega=5.12$ and a driving 
amplitude $A=0.5<A_{th}=2.05$. One can observe, similarly to the lower
Fig.~\ref{2brea}, that a moving discrete breather appears at the left boundary and
propagates inside the bulk, leaving behind the static solution.}
\label{pppp}\end{center}
\end{figure}

Here $x_0$ is defined as follows
\begin{equation}
x_0=\mbox{acos}(A/a)~, 
\label{initial}
\end{equation}
where $A$ is the driving amplitude. The breather frequency $\omega_B(a)$ depends 
on amplitude $a$ as follows
\begin{equation}
\omega_B(a)\simeq 1.03 \frac{\sqrt{3\pi^2(4+9a^2)}}{4K(s)}, \label{6}
\end{equation}
where $K(s)$ is the complete elliptic integral of the first kind with argument
$s=3a/\sqrt{2(9a^2+4)}$ and the factor $1.03$ takes into account a rescaling of 
the frequency of the ``tailed" breather~\cite{note} (see also \cite{kos}). 
As previously for the case of the envelope soliton solution, we perform a numerical
experiment where we put initially on the lattice the breather solution
of formula (\ref{breather}).  
Choosing the plus sign in this expression, we do not observe any significant 
transmission of energy inside the chain. Instead, the minus sign  
causes the appearance of a moving breather, which travels inside the chain leaving 
behind the static breather solution with plus sign. Fig.~\ref{pppp} presents 
this numerical experiment. 

The static breather solution (\ref{breather}) ceases to exist 
if the driving amplitude exceeds the threshold 
$A_{th}$ given by the resonance condition
\begin{equation}
\omega=\omega_B(A_{th}). \label{thre}
\end{equation}
Above this threshold the supratransmission process begins via the emission 
of a train of moving breathers from the boundary, exactly as it happens in the
case of small amplitudes. It should be mentioned again that the transmission regime 
is established due to moving discrete breathers. It has been remarked~\cite{kosprl} 
that discrete breathers are characterized by quantized velocities, while their frequency 
is given by the same formula (\ref{6}). This explains why one can use resonance 
condition (\ref{thre}) for the static discrete breather solution (\ref{breather}) 
to define the supratransmission threshold in the large amplitude limit.

\subsection{Supratransmission threshold: numerical test}

To check these predictions, we have performed a numerical determination of 
$A_{th}$ for several values of $\omega$, starting the chain at rest. This is 
accomplished by gradually increasing  $A$ and looking
for the minimal value $A_{th}$ for which a sizeable energy propagates into the
bulk of the chain. At early time, the scenario is qualitatively similar
to the one shown in Fig.\ref{transm}. Later on, the interaction of nonlinear
and quasi-linear modes and their ``scattering"  with the dissipating right boundary
establishes a steady energy flux into the chain. A conducting steady state, which 
is present also below $A_{th}$, will be discussed in Section V in connection with 
a lower-transmission threshold $A_{th}^-$.

\begin{figure}[t]
\begin{center}\leavevmode
\includegraphics[width=0.8\linewidth]{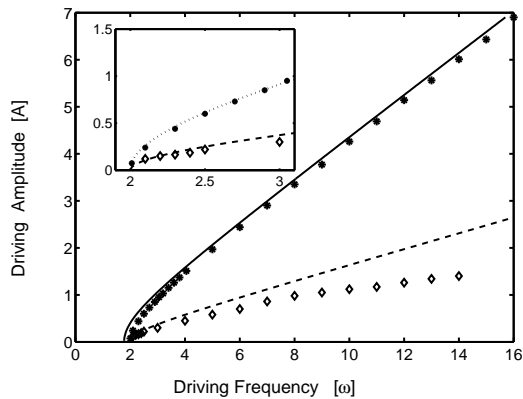}
\caption{
Comparison between analytic estimates and numerical values of threshold amplitudes
vs. the driving frequency. 
Main plot: the full dots are the numerical values 
of $A_{th}$ and the solid line is a plot of formulae (\ref{6}-\ref{thre}), which are 
valid for large amplitudes. The inset shows an enlargement of the small $A_{th}$ region, 
in order to illustrate the accuracy of the small-amplitude approximation (\ref{5}) (dotted 
line). The diamonds are simulation data for 
the lower-transmission threshold $A_{th}^-$ and the dashed line is formula (\ref{33}).
Notice how the latter is accurate only for small enough amplitudes (see
again the inset).
}
\label{raru}
\end{center}
\end{figure}

As seen in Fig.~\ref{raru}, formulae
(\ref{thre}) [with definition (\ref{6})] and (\ref{5}) (see the inset) are in excellent 
agreement with simulations for large  $A>2$ and small $A\lesssim 1$ amplitudes, 
respectively. The accuracy of the analytical 
estimate in formulae (\ref{thre}) and (\ref{5}) is of the order of few percents, 
at worst, in the intermediate amplitude range. We don't discuss here the lower curves 
in Fig.~\ref{raru}, which are related to the lower-transmission threshold. 

For comparison, we have checked that the supratransmission threshold is definitely
not associated with the quasi-harmonic waves with nonlinear
dispersion relation (\ref{2}). If this were the case,  the transmission should
start when the oscillation amplitude reaches the value for which the resonance condition 
$\omega=\omega_0(k,A)$ holds. As
$\omega_0(k,A)$ is maximal for $k=\pi$, we can get the expression
for the threshold value from the relation $\omega=\omega_0(\pi,A_{th})$, i.e.
\begin{equation} 
\omega^2=4+12A_{th}^2.  
\label{3} 
\end{equation}  
The amplitude values one obtains from Eq.~(\ref{3}) are far away
from the  numerical values and we don't even show them in Fig.~\ref{raru}. 
This is a further confirmation that supratransmission in the FPU model originates from 
direct discrete breather generation as it happens in the cases of discrete sine-Gordon 
and nonlinear Klein-Gordon lattices~\cite{leon}.

\section{Stationary states}

\begin{figure}[t]
\begin{center}
\includegraphics[width=0.9\linewidth,clip]{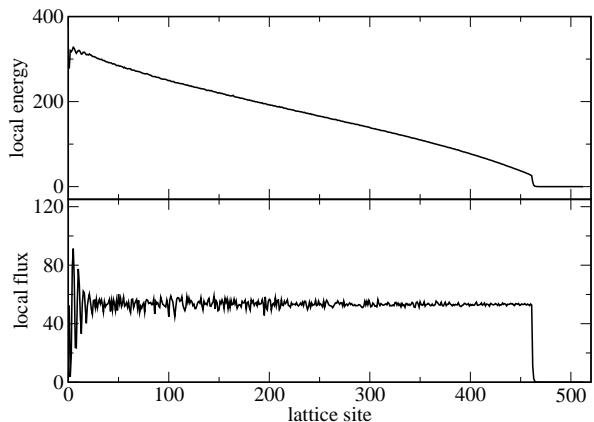}
\end{center}
\caption{
Time--averaged local energy (above) and energy flux (below) versus lattice position 
in the case of out-band driving, $\omega=3.5$, $A=1.27$. The 50 rightmost particle out of
$N=512$ have been damped with $\gamma=5$. The averages are taken over   
$2\, 10^5$ driving periods .}
\label{grad}
\end{figure}

As announced in the Introduction, we have also investigated the long-time
behavior of the chain.  As shown in the upper Fig.~\ref{grad} the time
averaged local energy (see formula (\ref{localenergy})) reaches asymptotically a
given profile: local energy monotonously decreases along the chain as in the case 
of simulations of stationary heat transport with two thermal baths~\cite{rep}. 
The time--average of the flux (\ref{flux}) in the stationary state is almost constant 
along the chain, apart from statistical fluctuations and some persistent flux oscillations 
at the left boundary. 

However, as we mentioned above, the value of the stationary flux
depends on the initial state of the chain. To illustrate this effect, let us excite 
the chain imposing a different boundary condition
\begin{equation}
u_0=\cos(\omega t)\left[B(1-e^{-t/\tau_1})+(A-B)(1-e^{-t/\tau_2})\right] 
\label{dr2} 
\end{equation}
where $\tau_2\gg\tau_1$ (in the experiment $\tau_2=10\tau_1=100$), 
$A<A_{th}$ and $B>A_{th}$.
Obviously, both the boundary condition (\ref{dr1}) and (\ref{dr2}) lead to the same 
driving amplitude $A$ for $t\gg \tau_2$.  However, at variance with (\ref{dr1}), when imposing 
(\ref{dr2}), the istantaneous forcing amplitude overcomes the critical amplitude $A_{th}$ 
for a time of the order of $\tau_2$, which is enough to establish a 
stationary flux regime. This drastically reduces the transmission threshold
to a value $A_{th}^-<A_{th}$, which we denote as {\it lower-transmission
threshold}. This is the first observation of this phenomenon, of which we will
give a theoretical interpretation in the following. The numerical determination
of $A_{th}^-$ versus the driving frequency $\omega$ is reported with diamonds
in  Fig.~\ref{raru}.

In the amplitude interval $[A_{th}^-,A_{th}]$, two steady states coexist, 
a {\it conducting} state and an {\it insulating} one. Each of the two steady
states can be attained with different initial conditions of the chain and
different driving pathways. For instance, the conducting state
is reached when imposing driving (\ref{dr2}), the insulating one when using
(\ref{dr1}). It is a typical {\it bistable} situation, where two (possibly chaotic)
attractors coexist in a given control parameter range.

\begin{figure}[t]
\begin{center}\leavevmode
\includegraphics[width=0.9\linewidth]{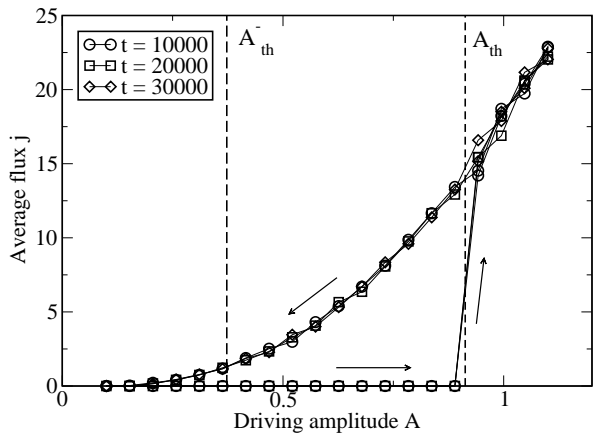}
\caption{
Energy flux versus driving amplitude for 
$\omega=3$, $N=512$, $\gamma=5$. The coexistence of the {\it conducting}
and {\it insulating} regimes is revealed by sweeping the forcing amplitude
in the range $[A_{th}^-,A_{th}]$. The sweeping direction is indicated by the
arrows. The analytical values of the two thresholds are indicated 
by the vertical dashed lines. While the prediction for the supratransmission threshold $A_{th}$ 
is quite good, the one for the lower-transmission threshold $A_{th}^-$ overestimates 
the numerical value (the prediction becomes better for smaller driving frequencies, as shown 
in Fig.~\ref{raru}). In order to show that the steady state is already reached for these
integration times, results for increasing averaging times are displayed with different
symbols.
}
\label{hyst2}
\end{center}
\end{figure}

This behaviour is illustrated in Fig.\ref{hyst2} using a different simulation
method. The average flux is computed after changing $A$ stepwise. A back and forth 
sweep around the amplitude interval $[A_{th}^-,A_{th}]$ reveals the presence
of the two states.

A justification of the presence of the lower-trasmission threshold can
be given in terms of quasi-linear theory. This theory leads to dispersion relation
(\ref{2}) only if one restricts to a single right-propagating mode.
However, due to reflection with the boundary and to mode interaction,
both the right-propagating mode and the left-propagating one can contribute
to the dispersion relation. In the Appendix, we derive this more general
dispersion relation. After introducing the complex mode amplitude $a_k$ for
the $k$-th mode, the dispersion relation takes the following form 
\begin{equation}
\omega(k)^2=2(1-\cos k)+3(1-\cos k)^2\left[|a_k|^2+2|a_{-k}|^2\right]~. 
\label{dispmin}
\end{equation}
In order to fulfill the resonance condition with both the right-propagating
($a_k$)  and the left-propagating ($a_{-k}$) mode, their amplitudes must be
equal $|a_k|=|a_{-k}|$. Since $\omega(k)$ is maximal for $k=\pi$, the condition
for the threshold amplitude is 
\begin{equation}
\omega^2 \;= \; 4+36\left(A_{th}^-\right)^2. 
\label{33}
\end{equation} 
This analytical estimate (dashed line in Fig.~\ref{raru}) fits well 
the numerical data only for driving frequencies close to the band edge (see the
inset). This can be justified by taking into account that dispersion relation (\ref{dispmin}) 
is valid only in the weakly nonlinear regime, i.e. mode amplitudes $|a_k|$ and $|a_{-k}|$  
much smaller than $1$. This condition certainly applies to the case in which the driving 
frequency is close to the band edge, since, then, the threshold 
amplitude $A_{th}^-$ is small. When the driving frequency is far from 
the band edge, one has to take into account higher-order corrections. 
The inclusion of the first ``satellite" mode ($3k$) produces a lower threshold
amplitude, but the agreement with numerical data extends only to slightly
larger amplitudes. To obtain a definitely better agreement, one should treat
all satellite modes $5k$, $7k$, etc.. We briefly discuss this aspect 
in the Appendix.

From the above considerations, it follows that the bistable nature of the energy flux can 
be explained making reference to the different excitations of the system. Indeed, with the 
system initially at rest, when following the driving method (\ref{dr1}), extended 
quasi-harmonic waves cannot be excited. Then, energy flow appears
only when the driving amplitude reaches the value necessary for localized mode
excitation. On the other hand, with driving (\ref{dr2}), the energy flow is
initiated by the overcoming of the supratransmission threshold and then sustained
also by extended quasi-harmonic waves.

It is also possible to give a heuristic argument to explain why the transition
from non zero to zero flux is ``continuous" at the lower-transmission threshold
$A_{th}^-$, while there is flux jump at the supratransmission threshold $A_{th}$.
When the quasi-harmonic waves are already excited, reducing the driving amplitude
diminishes also the number of resonating modes continuously. Hence, the flux goes
continuously to zero proportionally to this number, producing a sort of {\it second-order
phase transition}, when the flux is considered as an {\it order parameter}.
On the contrary, when increasing the driving amplitude with the lattice at rest 
across the supratransmission threshold  $A_{th}$, localized modes are excited, which 
successively excite also extended waves. Hence, a non zero flux is created suddenly 
from the zero flux state, generating a sort of {\it first-order phase transition}.

\section{Conclusions and perspectives}

We have discussed the supratransmission phenomenon for the Fermi-Pasta-Ulam
one-dimensional lattice. A theory, based on
a resonance condition of the driving frequency with the typical frequency of
localized excitations (solitons, breathers), gives a good agreement of the
supratransmission threshold with numerical data. Below this threshold two steady 
states coexist, a conducting and an insulating one. For even lower driving amplitudes a further
transition occurs to a region where only the insulating state persists: we have
called this new phenomenon lower-transmission threshold. Imposing a resonance
condition for nonlinear quasi-harmonic waves, we are able to derive an analytic 
expression for the lower-transmission threshold amplitude.

At the supratransmission threshold a jump in the energy flux appears. This is reminiscent
of a first-order phase transition. At variance, at the lower-transition threshold the
flux goes to zero continuously. This analogy with non-equilibrium phase
transitions~\cite{mukamel} should be further explored.

Fluctuations in steady states could be analysed to verify the possible role
played by the Gallavotti-Cohen out-of equilibrium fluctuation theorem~\cite{gallavotti}. 

The supratransmission phenomenon is quite generic and has already been observed
experimentally in a chain of coupled pendula~\cite{leon2}. Also the bistability of
conducting/insulating states is generic and could be observed experimentally in
similar conditions. For instance, one could apply this theory to micromechanical
experiments of the type performed by Sievers and coworkers~\cite{sievers}.

\acknowledgements
We thank J. Leon and D. Mukamel for useful discussions. This work is funded by the
contract COFIN03 of the Italian MIUR {\it Order and chaos in
nonlinear extended systems} and by the INFM-PAIS project {\it Transport
phenomena in low-dimensional structures}. One of the authors (R.K.) is also supported by the 
CNR-NATO senior fellowship 217.35 S and the USA CRDF award No GP2-2311-TB-02.

\appendix

\section{Nonlinear phonon dispersion}

In order to derive the nonlinear dispersion relation for extended quasi-harmonic 
waves, let us seek for the solutions of the equations of motion (\ref{1}) of the form
\begin{equation}
u_n=\frac{1}{2}\sum\limits_p \left[a_pe^{i(\omega(p) t+pn)}+a_{-p}^+e^{-i(\omega(p) t-pn)}
\right]~, 
\label{extended}
\end{equation}
where $\omega(p)$ is the frequency of the $p$-th mode and $a_p$ its complex
amplitude.
Substituting this Fourier expansion into the equations of motion, one gets the following
infinite set of algebraic equations for mode amplitudes~\cite{raleru}
\begin{equation}
\left[\omega(p)^2-2(1-\cos p)\right]a_p=
6\sum\limits_{q_1,q_2}G_{q_1,q_2}^p a_{q_1}a_{q_2}a_{q_1+q_2-p}^+,
\label{IIIeqa}
\end{equation}
where 
$$
G_{q_1,q_2}^p=\frac{1}{4} [1+\cos (q_1+q_2)+\cos (p-q_2)
+\cos (p-q_1)
$$
$$
-\cos p-\cos q_1-\cos q_2
-\cos (p-q_1-q_2) ].
$$

If only a single mode $p=k$ is excited, one gets the following dispersion relation
\begin{equation}
\omega(k)^2=2(1-\cos k)+3(1-\cos k)^2|a_k|^2, \label{disp1}
\end{equation}
which has been introduced in Eq.~(\ref{2}).

On the other hand, when both mode $k$ and mode $-k$ are excited, one obtains
\begin{equation}
\omega(k)^2=2(1-\cos k)+3(1-\cos k)^2\left[|a_k|^2+2|a_{-k}|^2\right]~, 
\label{disp2}
\end{equation}
which is presented as Eq.~(\ref{dispmin}) in the text.

As also mentioned in the text, one must sometimes consider the excitation of
``satellite" modes  $3k$, $5k$, etc.. The inclusion of the $3k$ mode produces
the addition of the following term
\begin{equation}
3\left[3\cos^2k-1-2\cos^3k\right]\left(|a_{-k}|^4+2|a_{-k}|^2|a_{k}|^2\right),
\end{equation}
to the r.h.s of Eq.~(\ref{disp2}).
This gives the following resonance condition at $k=\pi$
$$
\omega^2=4+36(A_{th}^-)^2\left(1+\frac{12(A_{th}^-)^2}{\omega^2-4}\right)~,
$$
where $\omega$ and $A_{th}^-$ are the driving frequency and lower-threshold amplitude,
respectively. Since the coefficient of the $(A_{th}^-)^4$ term in this relation is always
positive, the threshold amplitude one obtains is smaller that the one derived
from Eq.~(\ref{33}) in the text.

\medskip


\begin{thebibliography}{99}
\bibitem{note} Notice that a simpler approximate expression for the breather frequency 
has been proposed in~\cite{kosprl} in the form $\omega_B=\sqrt{3+81A^2/16}$. 
We have checked that this expression is also in good agreement with numerical data, 
but we prefer to use the more accurate form in formula (\ref{6}).
\end{thebibliography}
\end{document}